\documentclass[prd,showpacs,twocolumn,
amsmath,amssymb,superscriptaddress,floatfix,nofootinbib]{revtex4-1}
\usepackage[utf8]{inputenc}
\usepackage[sort&compress]{natbib}
\usepackage{ulem}
\usepackage{overpic}
\usepackage{bm}
\usepackage{times}
\usepackage{amssymb,amsbsy,amsmath,amsfonts}
\usepackage{graphicx}
\usepackage{float}
\usepackage{color}
\usepackage{booktabs}
\usepackage{makecell}
\usepackage{rotating}
\usepackage{srcltx}
\usepackage{slashed}
\usepackage{subfigure}
\usepackage{multirow}
\usepackage{verbatim}
\usepackage{hyperref}
\usepackage{tabularx}
\usepackage{lipsum} 

\DeclareUnicodeCharacter{3002}{HEREHEREHERE}

\begin{document}

\title{ Probing the structure of the $D_{s 0}^*(2317)$ and $X(3872)$ states through correlation functions }

\author{Yi-bo Shen}
\affiliation{School of Physics,  Beihang University, Beijing 102206, China}

\author{Zhi-Wei Liu}
\affiliation{School of Physics,  Beihang University, Beijing 102206, China}

\author{Jun-Xu Lu}
\affiliation{School of Physics,  Beihang University, Beijing 102206, China}

\author{Ming-Zhu Liu}
\email[Corresponding author: ]{liumz@lzu.edu.cn}
\affiliation{
Frontiers Science Center for Rare isotopes, Lanzhou University,
Lanzhou 730000, China}
\affiliation{ School of Nuclear Science and Technology, Lanzhou University, Lanzhou 730000, China}

\author{Li-Sheng Geng}\email[Corresponding author: ]{lisheng.geng@buaa.edu.cn}
\affiliation{School of Physics,  Beihang University, Beijing 102206, China}
\affiliation{Sino-French Carbon Neutrality Research Center, \'Ecole Centrale de P\'ekin/School of General Engineering, Beihang University, Beijing 100191, China}
\affiliation{Peng Huanwu Collaborative Center for Research and Education, Beihang University, Beijing 100191, China}
\affiliation{Beijing Key Laboratory of Advanced Nuclear Materials and Physics, Beihang University, Beijing 102206, China }
\affiliation{Southern Center for Nuclear-Science Theory (SCNT), Institute of Modern Physics, Chinese Academy of Sciences, Huizhou 516000, China}

\begin{abstract}

Over the past 20 years, many new hadron states have been discovered, but understanding their nature remains a key experimental and theoretical challenge.  Recent studies have established that hadron-hadron interactions primarily govern the generation of new hadronic states, with their spectroscopy serving as a powerful tool for probing these interactions and determining the corresponding compositeness. In this work, we study four scenarios to determine the  $DK$ interaction by reproducing the mass of the $D_{s0}^*(2317)$, i.e., assuming the $D_{s0}^*(2317)$ as a $DK$ molecule, a mixture of a $DK$ molecule and a bare state, a $DK-D_s\eta$ molecule, and a mixture of a $DK-D_s\eta$ molecule and a bare state. Using the $D^{0}K^{+}$ interactions derived from these scenarios, we predict the $D^{0}K^{+}$ correlation functions.  Our results demonstrate that the lineshape of the $D^{0}K^{+}$ correlation function is sensitive to the admixture effects from the coupled-channel $D^+K^0$ and the bare state. Furthermore, we find that the $D^{0}K^{+}$ correlation function can probe the position of the bare state, if such a QCD bare state exists.  Using the shallow-bound state candidate $X(3872)$ as input, we study the $D^0\bar{D}^{*0}$ correlation functions. These functions are highly sensitive to short-range dynamics and bare-state admixtures, resulting in clearly distinguishable correlation-function line shapes across different values of compositeness.

\end{abstract}
\date{\today}


\maketitle

\section{Introduction}

Quantum chromodynamics (QCD), the fundamental theory of the strong interaction, displays 
weak couplings at high energies, leading to asymptotic freedom. However, it reflects strong couplings at low energies, leading to color confinement, i.e., the
degrees of freedom become hadrons instead of quarks and gluons,
which challenges the study of the low-energy strong interactions
 at the quark level. The hadron spectroscopy is
essential for studying the strong interactions. A salient example is that the Cornell potential~\cite{Eichten:1974af}, proposed by studying the charmonium spectrum, verifies two fundamental properties of strong interactions in both the low-energy and high-energy regions. The Cornell potential has been confirmed by  Lattice QCD simulations~\cite{Bali:2000gf}. Later, fine interaction terms were incorporated into the Cornell potential, i.e., the Goldfrey-Isgur (GI) model~\cite{Godfrey:1985xj}, which explains most hadrons discovered by experiments at that time~\cite {Godfrey:1985xj,Capstick:1986ter}.

Hadrons are classified as mesons made of a quark and an anti-quark, and baryons made of three quarks in the conventional quark model~\cite{Gell-Mann:1964ewy}. Since 2003, an increasing number of states beyond the conventional quark model have been discovered~\cite{Brambilla:2010cs,Olsen:2017bmm,Brambilla:2019esw}, providing new insights into hadron structure and the underlying dynamics of strong interactions. Despite intensive experimental and theoretical studies~\cite{Chen:2016qju,Lebed:2016hpi,Oset:2016lyh,Esposito:2016noz,Dong:2017gaw,Guo:2017jvc,Ali:2017jda,Karliner:2017qhf,Guo:2019twa,Liu:2024uxn,Wang:2025sic,Doring:2025sgb}, the nature of these exotic states remains controversial, such as   $D_{s0}^{*}(2317)$~\cite{BaBar:2003oey} and $X(3872)$~\cite{Belle:2003nnu}.  
 Assuming  $D_{s_{0}}^{*}(2317) $ as an excited  $c\bar{s}$  meson, its mass is lower by around $160$ MeV than the prediction of the GI model~\cite{Godfrey:1985xj}.  The branching fraction $\mathcal{B}(D_{s0}^*(2317) \to \pi^0 D_s)$ was  determined to be approximately $1$ by the BESIII Collaboration~\cite{BESIII:2017vdm},  disfavoring  the excited  $c\bar{s}$ interpretation~\cite{Godfrey:2003kg,Colangelo:2003vg,Colangelo:2005hv}. Theoretical studies suggest instead that the $D_{s0}^{*}(2317)$ is likely a mixed state with a dominant molecular component~\cite{Ortega:2016mms,Albaladejo:2018mhb,Luo:2021dvj,Yang:2021tvc,MartinezTorres:2014kpc,Guo:2015dha,Yao:2015qia,Guo:2023wkv,Gil-Dominguez:2023huq}.   Similarly,    the precise analysis of the $X(3872)$ invariant  mass distributions  favors that $X(3872)$ contain a large molecular component~\cite{LHCb:2020xds,BESIII:2023hml}.

Studies of these exotic states have revealed that orbital excitation modes can be effectively replaced by multiple sea quark pairs, resulting in novel hadronic configurations. These configurations may form a distinct class of exotic states, i.e., hadronic molecules~\cite{Zou:2013af}. Therefore, hadron-hadron interactions, as a manifestation of non-perturbative QCD effects, play a fundamental role in the formation of such exotic states. This underscores the importance of precise hadron-hadron interactions as key inputs for elucidating the internal structure of exotic hadrons.
Fortunately, lattice QCD offers a first-principles and non-perturbative approach to investigate hadron-hadron interactions~\cite{Inoue:2010es,Lyu:2023xro,BaryonScatteringBaSc:2023zvt,Xing:2022ijm,Shi:2025ogt,Liu:2012zya,Mohler:2013rwa,Lang:2014yfa,Bali:2017pdv,Alexandrou:2019tmk}. Despite significant theoretical advancements in understanding these interactions, achieving precise determinations remains challenging due to the scarcity of experimental data. This highlights the urgent need for developing new experimental techniques to measure hadron-hadron interactions directly.

Recently, femtoscopy,  a technique that analyzes momentum correlations between particles emitted in high-energy collisions,  has emerged as a powerful alternative for probing the strong interaction~\cite{Fabbietti:2020bfg,Liu:2024uxn,Liu:2025rci}.  By measuring momentum correlation functions (CFs), femtoscopy has provided valuable insights into the interactions between hadrons.  This method exhibits an advantage in investigating hadron-hadron interactions involving unstable hadrons that are inaccessible through conventional scattering experiments~\cite{STAR:2014dcy,STAR:2015kha,ALICE:2019gcn,ALICE:2019hdt,ALICE:2020mfd,ALICE:2021cpv}, which motivates significant theoretical developments ~\cite{Morita:2014kza,Morita:2016auo,Ohnishi:2016elb,Haidenbauer:2018jvl,Morita:2019rph,Kamiya:2019uiw,Ogata:2021mbo,Kamiya:2021hdb,Haidenbauer:2021zvr,Liu:2022nec,Liu:2023uly,Liu:2023wfo,Molina:2023oeu,Liu:2024nac,Ge:2025put}.  Very recently, we demonstrated that the femtoscopy technique can help discriminate bound states, virtual states, or resonant states near mass thresholds,  offering tremendous potential to probe into the nature of exotic states~\cite{Liu:2024nac}.   As argued in Ref.~\cite{Epelbaum:2025aan}, the off-shell ambiguity in strong interactions is generally associated with the short‑distance behavior of relative wave functions, which introduces theoretical uncertainties into calculations of CFs. However, off‑shell ambiguities may not be particularly significant in practice. Some quantitative studies have indeed shown that such off‑shell effects are relatively mild within realistic interaction models for the two‑body problem~\cite {Gobel:2025afq,Molina:2025lzw}.  Hence, off‑shell ambiguities certainly exist, but their impact on two‑body systems may not be very large.

 In general, the determination of hadron-hadron interactions enables the evaluation of a state's compositeness through the Weinberg compositeness criterion~\cite{Weinberg:1962hj,Weinberg:1965zz}. This establishes a direct relationship between the interaction properties and the composite nature of a state~\cite{Wu:2025fzx}.  CFs serve as a powerful tool to extract these hadron-hadron interactions and consequently can help reveal the underlying compositeness. However, the quantitative impact of compositeness on  CFs remains poorly understood.  
 In this work, we investigate the $D_{s0}^*(2317)$ in the following four distinct scenarios:   a pure $DK$ molecular state (Scenario I), a mixture of a $DK$ molecular state and a bare state (Scenario II),   a $DK$-$D_s\eta$ coupled-channel molecule (Scenario III), and a mixture of a $DK$-$D_s\eta$ molecular component and a bare $c\bar{s}$ excited state (Scenario IV). Using a model-independent effective field theory (EFT) approach~\cite{Wu:2025fzx}, we extract the $D^{0}K^{+}$ interaction potential and subsequently predict the corresponding  CFs for each scenario. Assuming $X(3872)$ as a predominantly  molecule, we further investigate the $D\bar{D}^*$ CFs  using the similar approach.

This work is organized as follows. In Sec .~II, we present the formulas for calculating  CFs, scattering length, and effective range through the $T$ scattering matrix.  The numerical results and 
  detailed discussions are provided  in
Sec.~III.  Finally,  a summary is given in the last section.

\section{Theoretical Formalism}

In this section, we outline the theoretical framework for calculating CFs. Based on the Koonin-Pratt formula~\cite{Koonin:1977fh,Pratt:1990zq}, the CFs are determined by two key ingredients:    1) Particle-Emitting Source: Characterizing the spatial distribution of hadron emissions in relativistic heavy-ion collisions; 2) Scattering Wave Function: Encoding the final-state interactions between hadron pairs and derived from the reaction amplitude $T$-matrix. The latter provides direct access to the hadron-hadron interaction of the system under study.

First, we present  the CFs in the single channel case~\cite{Liu:2024nac},
\begin{align}
  C(k)&=1+\int\limits_0^\infty{\rm d}^3r~S_{12}(r)\nonumber\\
  &\times\left[\left|j_0(kr)+T(\sqrt{s})\cdot\widetilde{G}(r,\sqrt{s})\right|^2-|j_0(kr)|^2\right],\label{cf1}
\end{align}
where  $S_{12}$   is characterized by a  Gaussian source function  $S_{12}(r)=\text{Exp}~[-r^{2}/(4R^2)]/(2\sqrt{\pi}R)^3$ with a single parameter $R$,  $j_{0}(kr)$ is the spherical Bessel function, and  $k=\sqrt{s-(m_1+m_2)^2}\sqrt{s-(m_1-m_2)^2}/(2\sqrt{s})$ represents the center-of-mass (c.m.) momentum of the particle pair with the masses being  $m_1$ and $m_2$ and c.m. energy $\sqrt{s}$.  The $\widetilde{G}$ is given by
\begin{equation}
\widetilde{G}(r, \sqrt{s})=\int_0^{q_{\text {max }}} \frac{\mathrm{d}^3 k^{\prime}}{(2 \pi)^3} \frac{\omega_1+\omega_2}{2 \omega_1 \omega_2} \frac{j_0\left({k^{\prime}} r\right)}{s-\left(\omega_1+\omega_2\right)^2+i \varepsilon}
\end{equation}
with $\omega_{i}(k)=\sqrt{m_i^2+k^{2}}$$(i=1,2)$.
$\widetilde{G}$ is regularized by a sharp cutoff $q_{\rm max}$ of the order of $0.5-1.5$ GeV~\cite{Liu:2025oar}.

The $T(\sqrt{s})$ matrix is obtained by solving the  Lippmann-Schwinger equation
\begin{equation}
    T(\sqrt{s})=\frac{V(\sqrt{s})}{1-V(\sqrt{s})\cdot G(\sqrt{s})},
\label{T}
\end{equation}
where $G(\sqrt{s})$ is the loop function
\begin{equation}
\begin{split}
    G(\sqrt{s})&=\int_{0}^{q_\mathrm{max}}\frac{{\rm d}q~q^{2}}{4\pi^{2}}\frac{\omega(m_{1},q)+\omega(m_{2},q)}{\omega(m_{1},q)\cdot\omega(m_{2},q)}\\
    &\times\frac{1}{s-[\omega(m_{1},q)+\omega(m_{2},q)]^{2}+i\epsilon}.
\end{split}
\label{loop function}
\end{equation}

For the single-channel case, the $D_{s_{0}}^{*}(2317)$ is treated as a $DK$ bound state, and we parameterize the $DK$ interaction by the contact-range EFT,
\begin{equation}
    V=C_{a}. 
\label{V}
\end{equation}
To quantitatively assess the dressed contribution of a bare $c\bar{s}$ state to the $DK$ interaction potential, we incorporate an effective potential~\cite{Shi:2024llv},
\begin{equation}
V=\frac{\alpha}{\sqrt{s}-m_{bare}}, 
\label{v2}
\end{equation}
where $m_{bare}$ is the mass of the bare state predicted by the GI model and  $\alpha$ is an unknown parameter.

The above potentials can be determined by reproducing the mass and compositeness of the  $D_{s_{0}}^{*}(2317)$.  After   the $T$ matrix is determined, one can obtain the scattering length $a$ and effective range $r_0$  at the threshold~\cite{Ikeno:2023ojl},
\begin{equation}
    -\frac{1}{a}=-\left.8 \pi \sqrt{s} T^{-1}\right|_{s=s_{\mathrm{th}}},
\end{equation}
\begin{equation}
\begin{aligned}
r_{0} & =\frac{\partial}{\partial k^{2}} 2\left(-8 \pi \sqrt{s} T^{-1}+i k\right) \\
& =\left.\frac{\sqrt{s}}{\mu} \frac{\partial}{\partial s} 2\left(-8 \pi \sqrt{s} T^{-1}+i k\right)\right|_{s=s_{\text {th }}},
\end{aligned}
\end{equation}
where $s_\mathrm{th}=(m_1+m_2)^2$.

The compositeness of a physical state is determined by the following expression~\footnote{ We note an extra factor of $2\sqrt{s}$ relative to the definition of the function of $s$ in Ref.~\cite{Li:2024tof}.  }
\begin{equation}
    P = -\lim_{\sqrt{s}\to \sqrt{s_0}}\left(\sqrt{s}-\sqrt{s_0}\right)T(\sqrt{s})\frac{\partial~G(\sqrt{s})}{\partial \sqrt{s}}.  
    \label{X2}
\end{equation}

Next, we present the CFs in the coupled-channel case
\begin{equation}
\begin{aligned}
\label{kpcouplechannel}
& C\left(k\right)=1+4 \pi \theta\left(q_{\max }-k\right) \times \\
& \int_0^{+\infty} d r r^2 S_{12}(r)\left\{\left|j_0\left(k r\right)+T_{11}(\sqrt{s}) \widetilde{G}^{(1)}(r, \sqrt{s})\right|^2\right. \\
& \left.+\left|T_{21}(\sqrt{s}) \widetilde{G}^{(2)}(r, \sqrt{s})\right|^2-j_0^2\left(k r\right)\right\}.  
\end{aligned}
\end{equation}

For the coupled-channel case, the loop function in matrix form reads
\begin{equation}
    G=\left(\begin{array}{cc}
G_{DK} & 0 \\
0 & G_{D_{s}\eta}
\end{array}\right),
\end{equation}
and the coupled-channel potential in matrix form is written as~\cite{Guo:2006fu,Liu:2012zya,Altenbuchinger:2013vwa,Guo:2015dha}
\begin{equation}
    V_{DK-D_s\eta}^{J^P=0^+}=\begin{pmatrix} 
     C_a&  \frac{\sqrt{3}}{2} C_a\\ \frac{\sqrt{3}}{2}  C_a& 0\end{pmatrix}.
\label{v3}
\end{equation}
Taking into account the dressed effect of the bare state, the coupled-channel  potential becomes
\begin{align}
    V=\left(\begin{array}{cc}
C_a+\frac{\alpha A_iA_i}{\sqrt{s}-m_{bare}} & \frac{\sqrt{3}}{2}C_{a}+\frac{\alpha A_iA_j}{\sqrt{s}-m_{bare}}  \\
\frac{\sqrt{3}}{2}C_{a}{+\frac{\alpha A_iA_j}{\sqrt{s}-m_{bare}}}  & 0
\label{v4}
\end{array}\right), 
\end{align}
 where the coefficients $A_i$ ($i$ denotes channel $i$) are associated with the group representation~\cite{Lin:2025pyk}.  Following Ref.~\cite{Gamermann:2006nm}, we have the $A_{DK}=1$ and $A_{D_s\eta}=\frac{1}{\sqrt{3}}$.






After determining the $T$ matrix, the scattering lengths, effective ranges, and compositeness can be obtained by 
\begin{align}
    -\frac{1}{a_{1}} &= \left.(-8 \pi \sqrt{s})T_{11}^{-1}\right|_{s_{\mathrm{th} 1}},
    \label{a1}
\end{align}
\begin{align}
    r_{0,1}= & \left.2 \frac{\sqrt{s}}{\mu_{1}} \frac{\partial}{\partial s} ( - 8 \pi \sqrt { s } ) T_{11}^{-1}\right|_{\mathrm{s}_{\mathrm{th} 1}},
    \label{r1}
\end{align}
\begin{align}
    -\frac{1}{a_{2}} &= \left.(-8 \pi \sqrt{s})T_{22}^{-1}\right|_{s_{\mathrm{th} 2}},
    \label{a2}
\end{align}
\begin{align}
    r_{0,2}= & \left.2 \frac{\sqrt{s}}{\mu_{2}} \frac{\partial}{\partial s} ( - 8 \pi \sqrt { s } ) T_{22}^{-1}\right|_{\mathrm{s}_{\mathrm{th} 2}},
    \label{r2}
\end{align}
\begin{equation}
P_{i}=-g_{i}^2 \frac{\partial~G_{ii}(\sqrt{s})}{\partial \sqrt{s}},  
\end{equation}
where $g_{i}$ is the coupling between
the molecular state and its constituents from the residues of the corresponding pole,
\begin{equation}
g_{i}g_{j}=\lim_{\sqrt{s}\to \sqrt{s_0}}\left(\sqrt{s}-\sqrt{s_0}\right)T_{ij}(\sqrt{s}).
\end{equation}

\section{Results and discussions}

In this section, considering the exotic state $D_{s0}^{*}(2317)$ as a predominantly $DK$ molecular state, we extract the $DK$ potential and then present the results for the $DK$ scattering length and CFs in four scenarios. Using a similar strategy, we show the results for the $\bar{D}^*D$  scattering length, effective range,  and CFs by assuming $X(3872)$ as a shallow $\bar{D}^*D$ bound state. 
In addition to the potential parameters, the cutoff $q_{max}$ also needs to be determined, for which we choose a value of 1 GeV without losing generality.

\subsection{Single-channel case}

\begin{figure}
\includegraphics[width=0.45\textwidth]{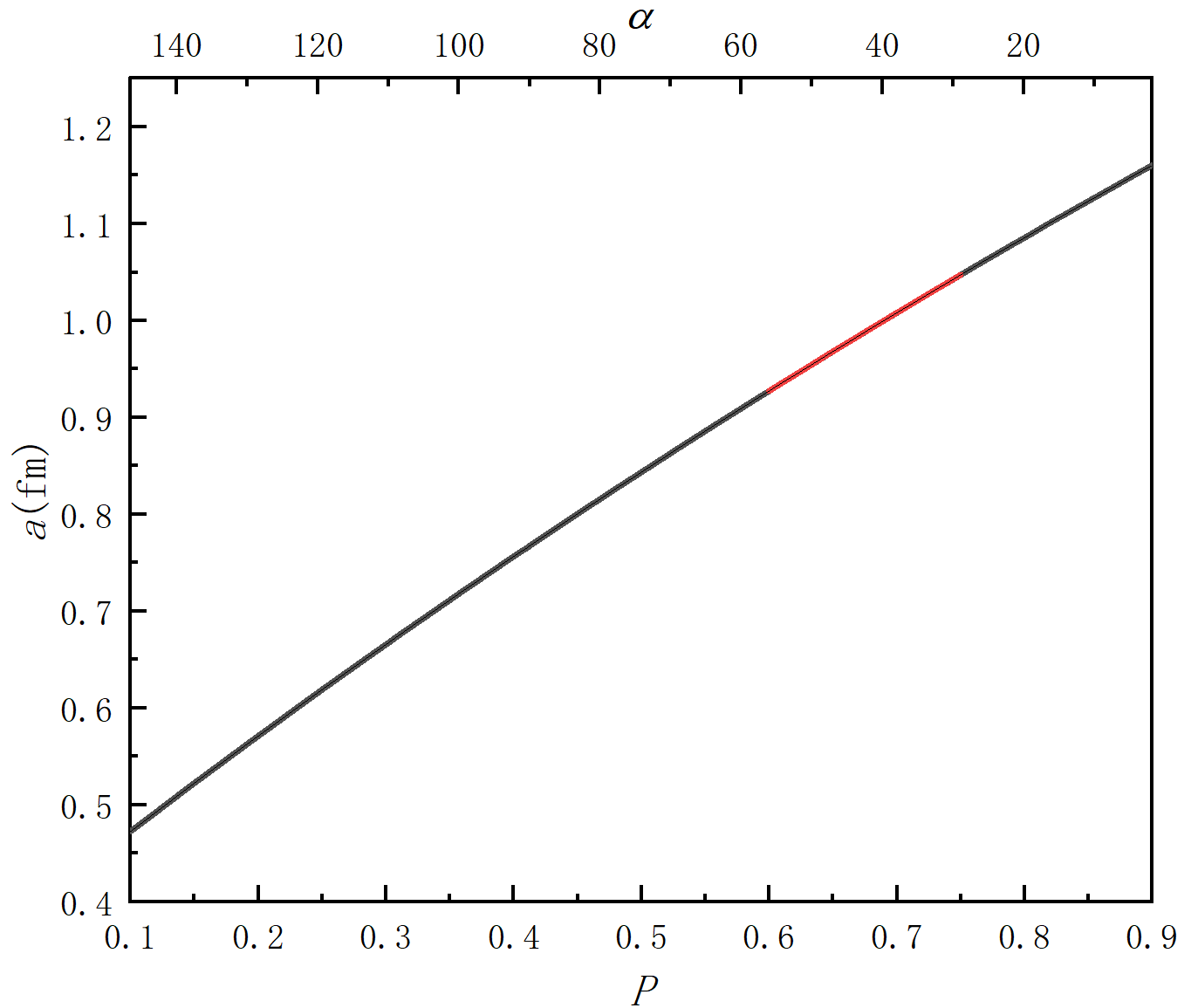} 
\caption{   $DK$ scattering length($I=0$) as a function of the compositeness for the $D_{s0}^*(2317)$( the bottom horizontal axis) and the parameter $\alpha$ (the top horizontal axis) in the single-channel case. \label{a varied with p}
}
\label{a}
\end{figure}

\begin{figure}
\includegraphics[width=0.45\textwidth]{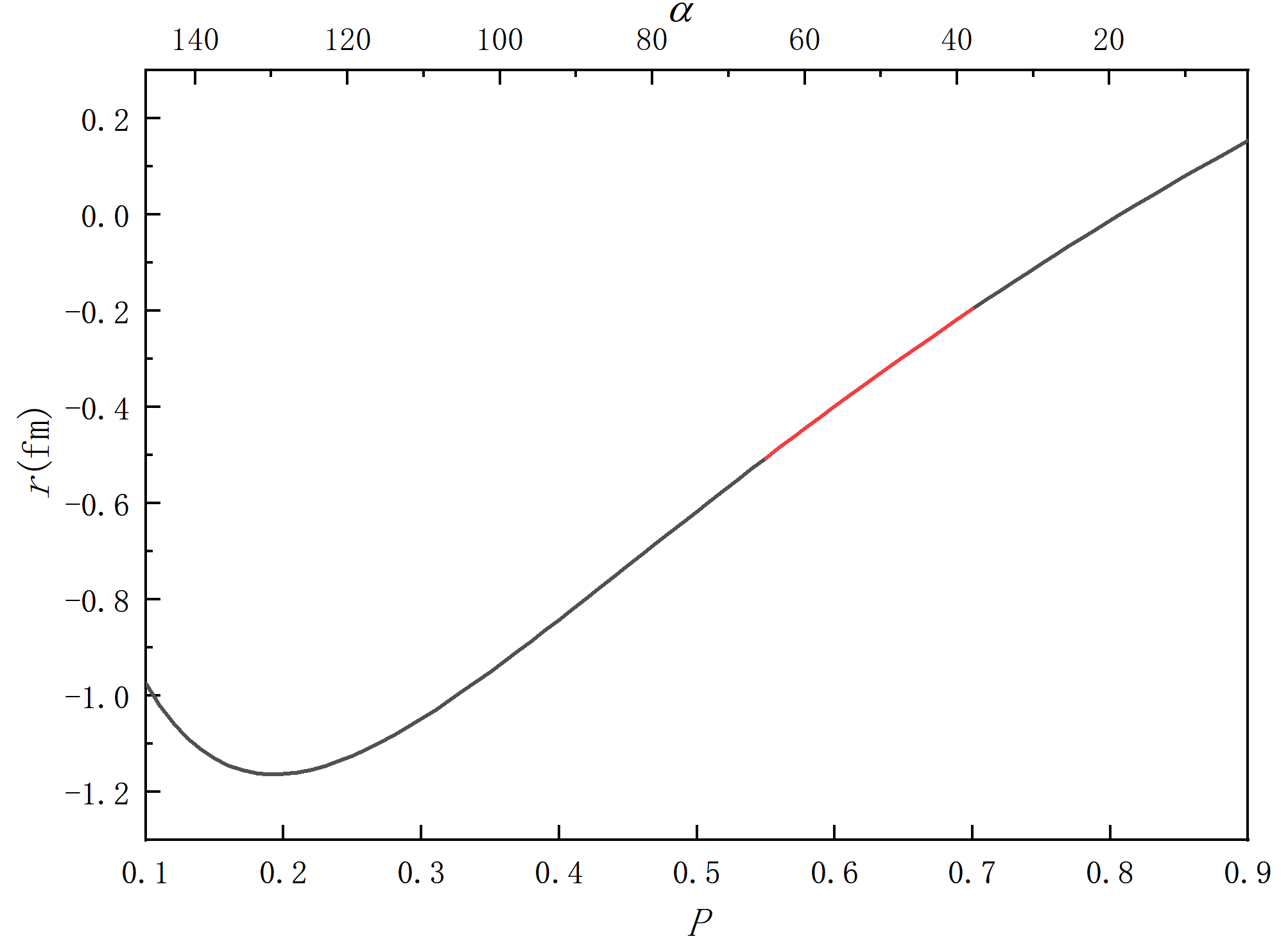} 
\caption{    $DK$ effective range($I=0$) as a function of the compositeness for the $D_{s0}^*(2317)$ (the bottom horizontal axis) and the parameter $\alpha$ (the top horizontal axis) in the single-channel case.    \label{r varied with p}
}
\label{r}
\end{figure}

We first focus on the single-channel case.   Since the  $DK$ potential in Scenario I  is a constant, we can determine its value by reproducing the mass of $D_{s0}^{*}(2317)$.  Our results indicate that the compositeness $P$ consistently equals unity, while the scattering length is determined to be $1.23$ fm. In Scenario II,  the potential has two parameters, which can be determined by reproducing the mass and compositeness of the $D_{s0}^{*}(2317)$. A lot of theoretical studies indicated that the molecular component accounts for around $70\%$ of the $D_{s0}^*(2317)$ wave function~\cite{MartinezTorres:2014kpc,Albaladejo:2018mhb,Yang:2021tvc,Guo:2023wkv,Gil-Dominguez:2023huq}. We find the scattering length $1.0$~fm in Scenario II with the compositeness $70\%$,  smaller than that in Scenario I.  
To further quantify the dressing effects of the bare state,  we plot the scattering length as a function of compositeness and $\alpha$  in Fig.~\ref{a}. The red line corresponds to a compositeness range of   $0.60\sim0.75$, yielding a scattering length of $0.92\sim1.05$ fm,  in agreement with the Lattice QCD calculations~\cite{Liu:2012zya}.    Our findings demonstrate that including a bare state significantly modifies the $DK$ scattering length.   The effective range as a function of compositeness and the parameter $\alpha$ is shown in Fig.~\ref{r varied with p}. As the compositeness of the $D_{s0}^{*}(2317)$ decreases, the effective range also decreases, eventually becoming negative for a compositeness of $P = 0.8$. These results indicate that the effective range is highly sensitive to the compositeness.    

\begin{figure}
\includegraphics[width=0.45\textwidth]{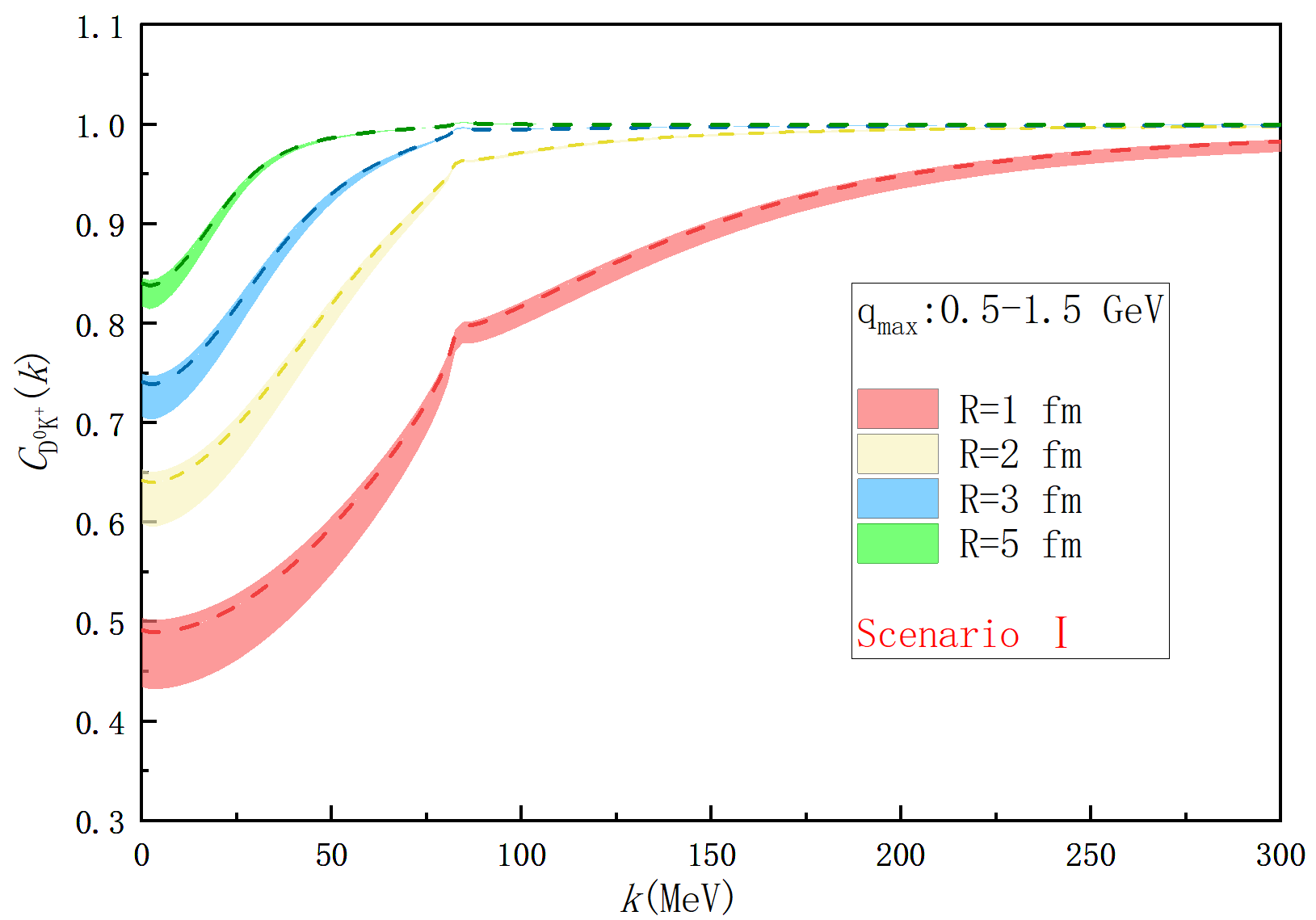} 
\caption{$D^{0}K^{+}$ CFs for $R=1, 2, 3, 5$ fm in the single-channel case.  \label{single channel case}
}
\end{figure}

\begin{figure}
\includegraphics[width=0.45\textwidth]{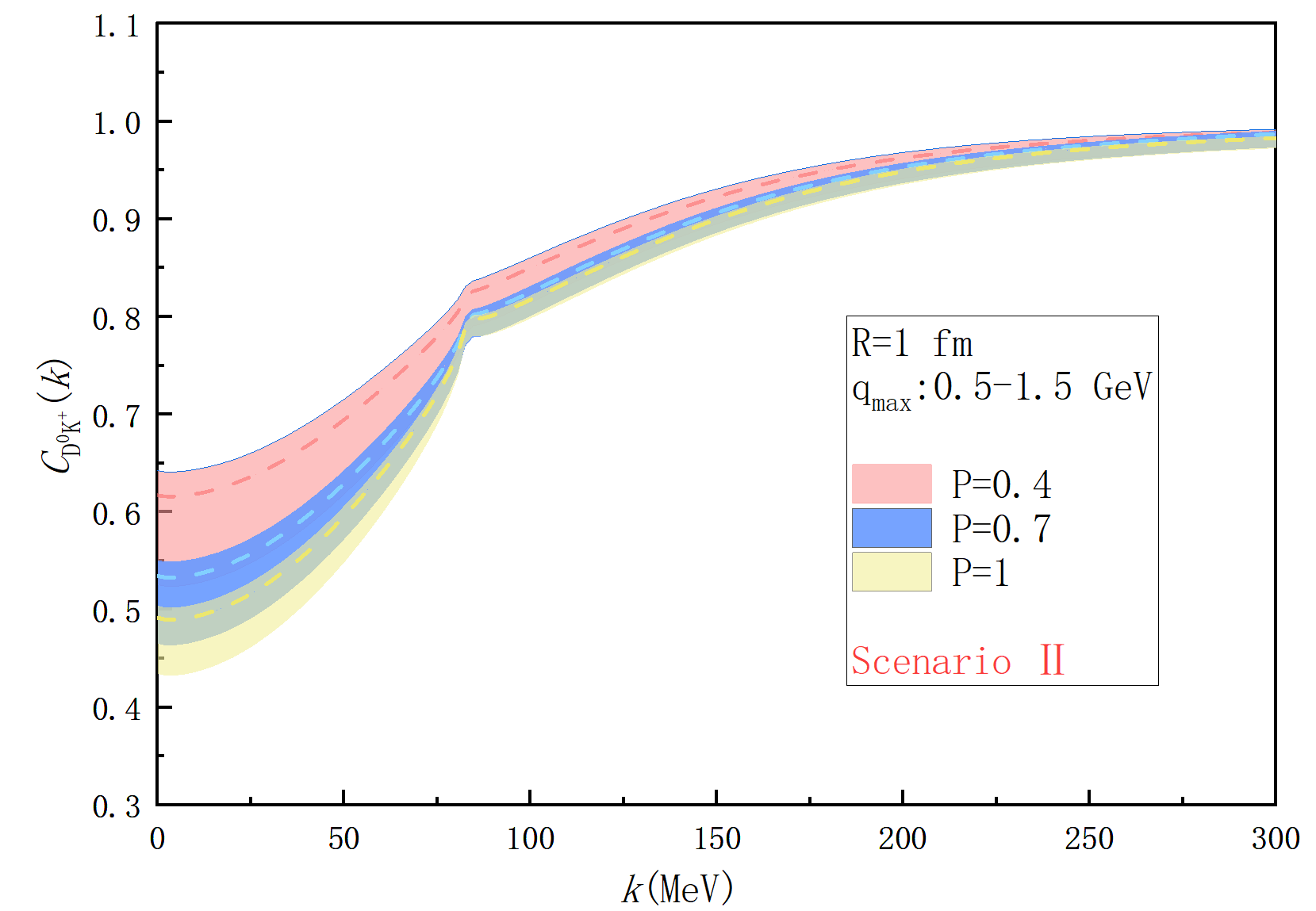} 
\caption{$D^{0}K^{+}$ CFs as a function of the compositeness for $P=0.4, 0.7$ and $1$ in the single-channel case dressed by a bare state. 
\label{single and bare}
}
\end{figure}

Once the $DK$  potential is obtained, we can calculate its corresponding  CFs as well. In Fig.~\ref{single channel case}, we present the  $D^{0}K^{+}$ CFs in Scenario I,   where the dashed lines correspond to a cutoff of $q_{max}=1$~GeV, and the bands represent the range of results obtained by varying  $q_{max}$   from $0.5$ to $1.5$ GeV.      
We can see that the $D^{0}K^{+}$ CFs are less than 1  for a strongly attractive interaction, consistent with the general feature of CFs~\cite{Liu:2023uly}. Furthermore, our analysis reveals an inverse relationship between the source size $R$ and the $D^{0}K^{+}$  CFs.   For $R = 1$ fm, the  $D^{0}K^{+}$ CFs exhibit the largest deviation from unity. We therefore select $R = 1$ fm for our subsequent analysis.   To investigate the relationship between the CFs and compositeness, we calculate the  $D^{0}K^{+}$ CFs in Scenario II,  as shown in Fig.~\ref{single and bare}. As the proportion of the $DK$  molecular configuration decreases, the corresponding $D^{0}K^{+}$ CF values approach unity, showing that including a bare state modifies the line shape of CFs.

\begin{figure}
\includegraphics[width=0.45\textwidth]{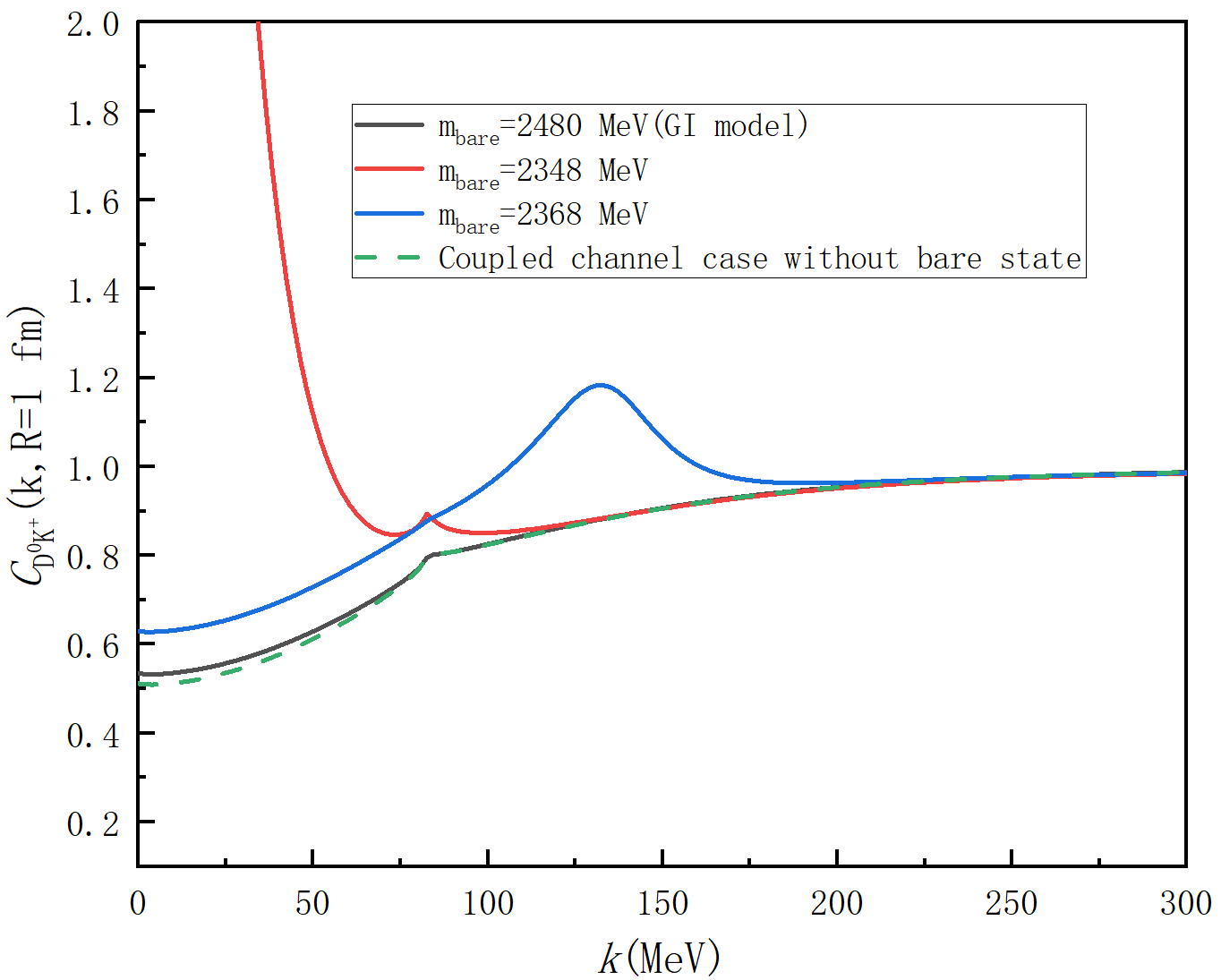} 
\caption{ $D^0K^+$ CFs for $P=0.7$ (solid lines) in scenario II but for different masses  of the  bare state.
}
\label{mass of bare}
\end{figure}

\begin{figure}
\includegraphics[width=0.45\textwidth]{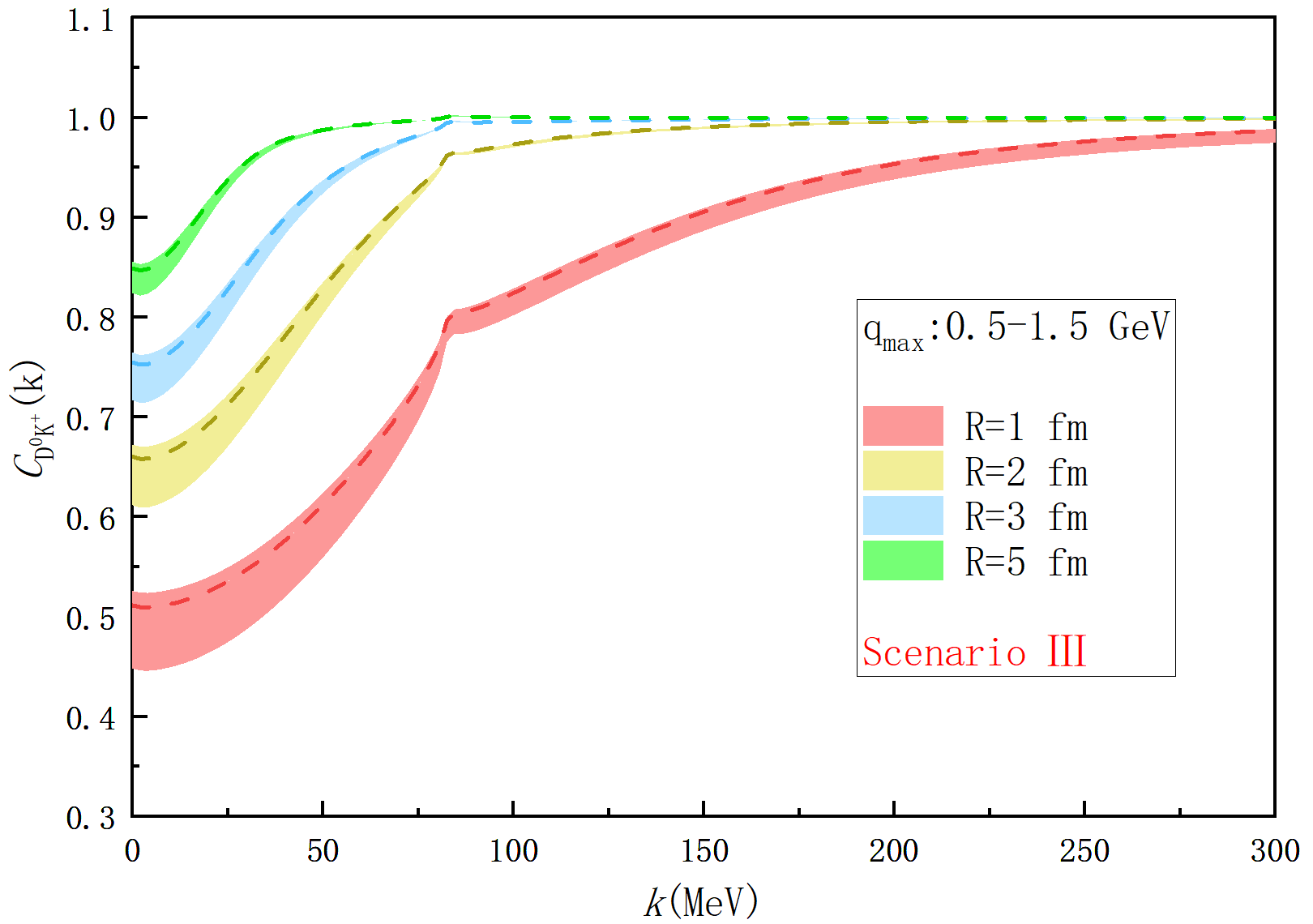} 
\caption{$D^{0}K^{+}$ CFs as a function of the compositeness for $R=1, 2, 3, 5$ fm in the coupled-channel case.  \label{coupled channel 1}
}
\label{couple}
\end{figure}

\begin{figure}
\includegraphics[width=0.45\textwidth]{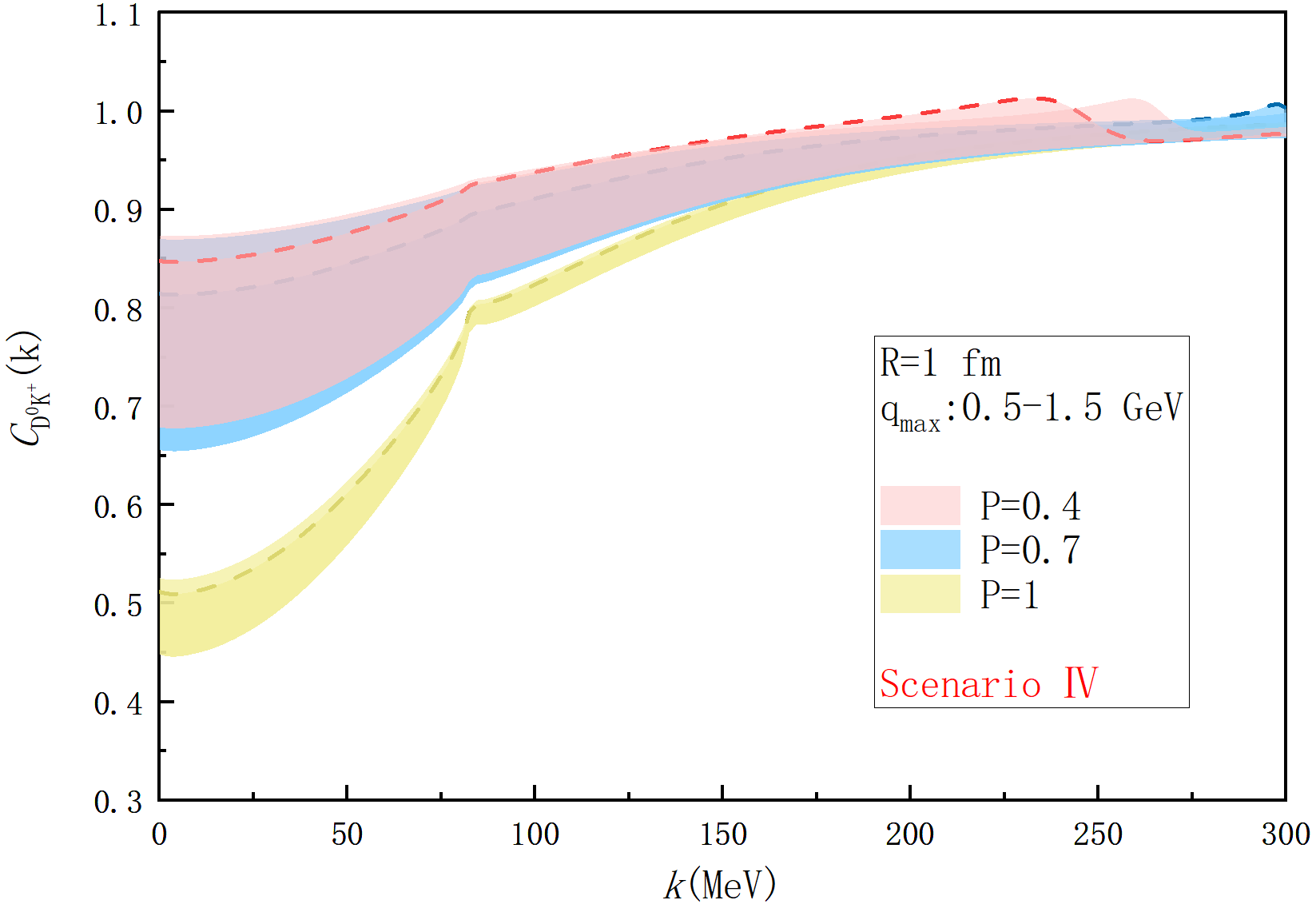} 
\caption{$D^{0}K^{+}$ CFs as functions of the compositeness for $P_{1}=0.4, 0.7$ and $1$ in the coupled-channel case dressed by a bare state.  \label{coupled channel}
}
\label{couple}
\end{figure}

 We further investigate the impact of the mass of the bare state on the CFs. The mass of the bare state is set to two different values:  $m_{bare}=2.348$~GeV(below the $DK$ threshold) and $m_{bare}=2.368$~GeV(above the threshold). The corresponding  CFs are shown in Fig.~7. It is evident that the mass of the bare state significantly affects the line shape of the CFs. Due to the singularity in the scattering matrix $T$, when the mass of the bare state is chosen near but above the threshold, a low-momentum peak emerges. Moreover, due to the continuity of the CFs, if the bare-state mass lies below the threshold, a peak near zero momentum appears, leading to enhanced CFs at low momentum. These distinct CF behaviors may provide a new avenue for studying how bare states affect hadron-hadron interactions, improving our understanding of the complex dynamics of strong interactions.

\subsection{Coupled-channel case}

We now extend our analysis to the coupled-channel case, employing the same theoretical framework previously applied to the single-channel case.  First, we analyse the 
 $DK$ scattering length  $a$. Our analysis yields   $a=1.17$ fm for  Scenario III (the compositeness $P_1 + P_2 = 1$), while for Scenario IV (the compositeness $P_1 + P_2 \approx 0.7$)  we obtain $a={1.01}$ fm.  We find that the results for Scenarios I and III, as well as Scenarios II and IV, are consistent with one another, which indicates that the coupled-channel effects affect the $DK$ scattering length slightly.  As for the cases with and without a bare state,  we find that the dressing effects of the bare state are pronounced for the   $DK$ scattering length.

We proceed to calculate the  $D^{0}K^{+}$ CFs in Scenario III, as presented in Fig.~\ref{coupled channel 1}. A comparison with the results of Scenario I  in Fig.~\ref{single channel case} reveals remarkably similar CF patterns, suggesting that the coupled-channel effects are relatively modest. Furthermore, we systematically examine the  $D^{0}K^{+}$  CFs in Scenario IV as functions of the compositeness in Fig.~\ref{couple}, analogous to the analysis in Fig.~\ref{single and bare}. 
When both coupled-channel and bare-state dressed effects are combined, the resulting difference between $P=1$ and $P=0.7$ becomes substantial, as illustrated in Fig.~6. The joint influence of the bare state and coupled-channel effects indeed induces this notable difference. It should be noted that such a coupled-channel effect is specific to $D^0K^+$ and $D^+K^0$ rather than $D_s\eta$. In other words, the impact of the $D_s\eta$ channel on the $DK$ CFs is negligible in Scenario II and Scenario IV, consistent with the scattering length analysis.

The coupled-channel effect manifests itself in the final-state interaction but is not incorporated into the emission source.
With the coupled-channel effect,   the CFs formula in Eq.~(\ref{kpcouplechannel})  becomes   
\begin{equation}
\begin{aligned}
C_j\left(k_j\right) & =1+\int_0^{\infty} \mathrm{d}^3 r S_{12}\left[\sum_i \omega_i\left|\delta_{i j} j_0\left(k_j r\right)+T_{i j} \widetilde{G}_i\right|^2\right. \\
& \left.-\left|j_0\left(k_j r\right)\right|^2\right],
\label{weight1}
\end{aligned}
\end{equation}
where the weight of each component is characterized by $\omega_i$. In this work, we estimate the   ratio between  different   weights  $\omega_{i}$ by~\cite{Liu:2024nac}
\begin{equation}
\label{temperature}
\frac{\omega_i}{\omega_j} \approx \frac{\exp \left[\left(-m_{i 1}-m_{i 2}\right) / T^*\right]}{\exp \left[\left(-m_{j 1}-m_{j 2}\right) / T^*\right]},
\end{equation}
where the hadronization temperature is $T^{*} = 154$~MeV.

\begin{figure}
\includegraphics[width=0.45\textwidth]{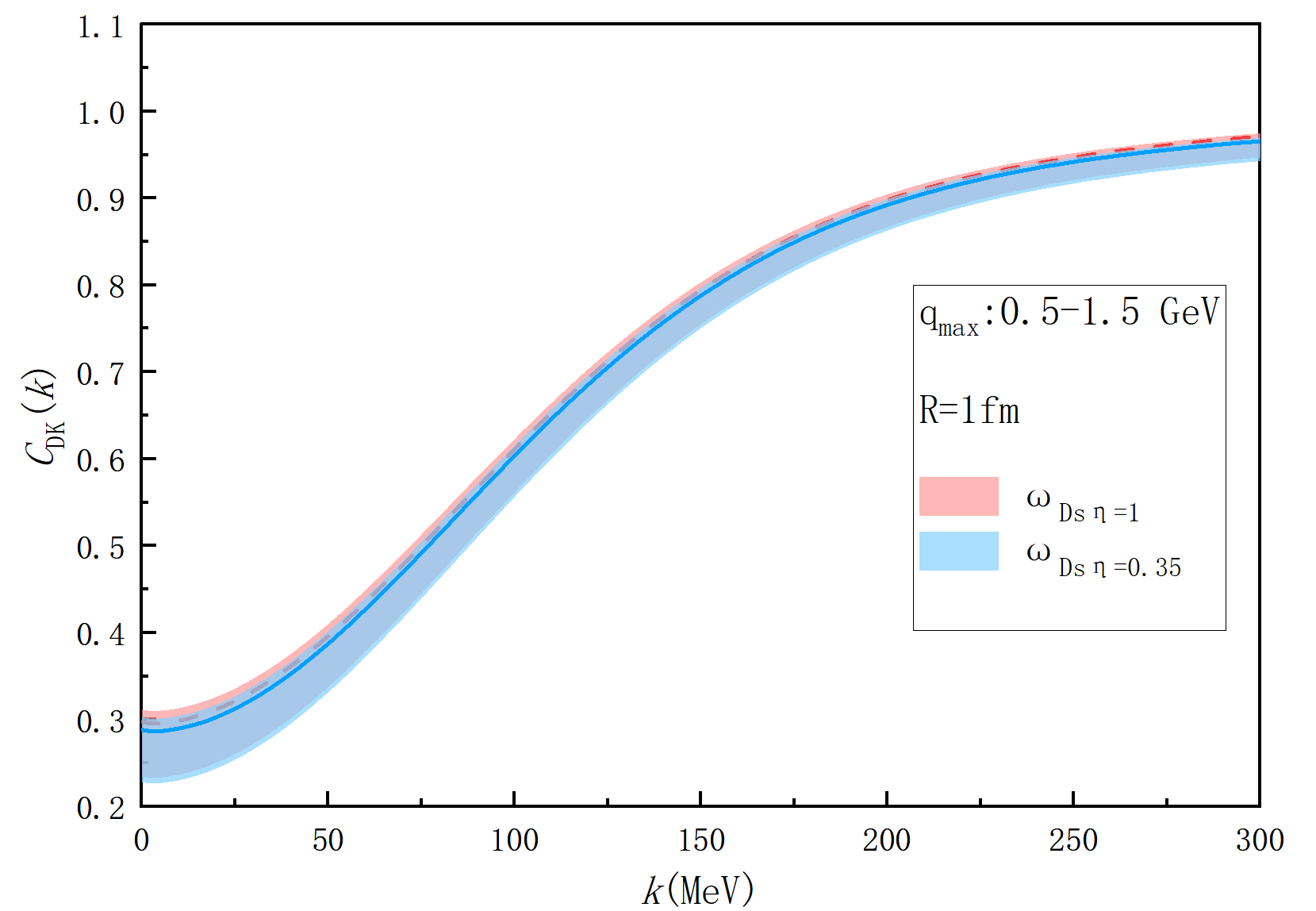} 
\caption{$DK$ CFs calculated with Eq.~\eqref{v3} and Eq.~\eqref{weight1} for  $R=1$ fm and different weights $\omega_{i}$. 
\label{weight}
}
\end{figure}

In Scenarios III and IV, we assume that the weights of the $DK$ channel and the $D_s\eta$ channel are the same.  Using Eq.~(\ref{temperature}), we obtain the ratio  $\omega_{DK}/\omega_{D_s\eta}=1/0.35$. Then we recalculate the $DK$ CFs as shown in Fig.~\ref{weight}, and the result remains almost unchanged, indicating that the coupled-channel effect is weakly dependent on the source function.  

  We observe that the uncertainty of CFs is larger at lower k than at higher k, and the uncertainty bands vary significantly with compositeness. This is because for any finite cutoff, the correlation function at high momentum always approaches unity. In contrast, at low $k$, the wave function exhibits little oscillation, so the result of the $r$-integration becomes sensitive to the cutoff. When the compositeness is close to unity, the state is predominantly molecular, and the low-energy scattering amplitude is largely governed by long-range dynamics, leading to a weak cutoff dependence and a narrow uncertainty band in the correlation function. In contrast, for smaller compositeness, the elementary component introduces strong short-range, energy-dependent interactions, making the low-momentum correlation function much more sensitive to the cutoff and thus leading to a significantly broader band.



\subsection{Inverse problem}

If the experimental  $DK$ CFs are measured, it would be intriguing to investigate whether the compositeness can be extracted from the  $DK$ CFs. To determine the potential parameters, we employ a random sampling method to select multiple points from the  $DK$ CFs in Scenario IV, treating them as synthetic experimental data:  According to the results of Scenario IV, we take a point about every $2$ MeV at at momentum less than 40, and we randomly obtain an error within ten percent of correlation functions, then we fit these data with Eq.(10) to fix the parameter in the potential. Ultimately, we obtained errors originating from the sampling points.   In Scenario IV, the potential is characterized by two parameters: $C_{a}=-95.32 $ and $\alpha=8.56$ GeV, and the corresponding compositeness is  $P_{1}$ = $0.7$.    By fitting the resampled data points, we obtain the potential parameters $C_a=-95.37\pm 15.65$ , $\alpha=8.41\pm 0.72$ GeV and  We calculated the compositeness $P_1$ using uncertainty propagation based on the $1-\sigma$ uncertainties of the potential parameters, yielding a result of $0.698 \pm0.06 $,  which are in excellent agreement with the original input values.

\subsection{$X(3872)$}
The $D_{s0}^*(2317)$ can be regarded as a deeply bound $DK$ molecule with a binding energy of $45$~MeV. 
In the following, we examine a shallow bound state, using the $X(3872)$ as a representative example. The hadron-hadron potential for $X(3872)$ is similar to those for the $D_{s0}^*(2317)$. Assuming  the $X(3872)$ as a mixture of a $\bar{D}^*D$ molecular state and a bare $c\bar{c}$ state, we can characterize  its potential as $C_a+\frac{\alpha}{\sqrt{s}-m_{c\bar{c}}}$, where the mass of the bare state is  $m_{c\bar{c}}=3950$ MeV taken from the GI model~\cite{Godfrey:1985xj}.

\begin{figure}
\includegraphics[width=0.45\textwidth]{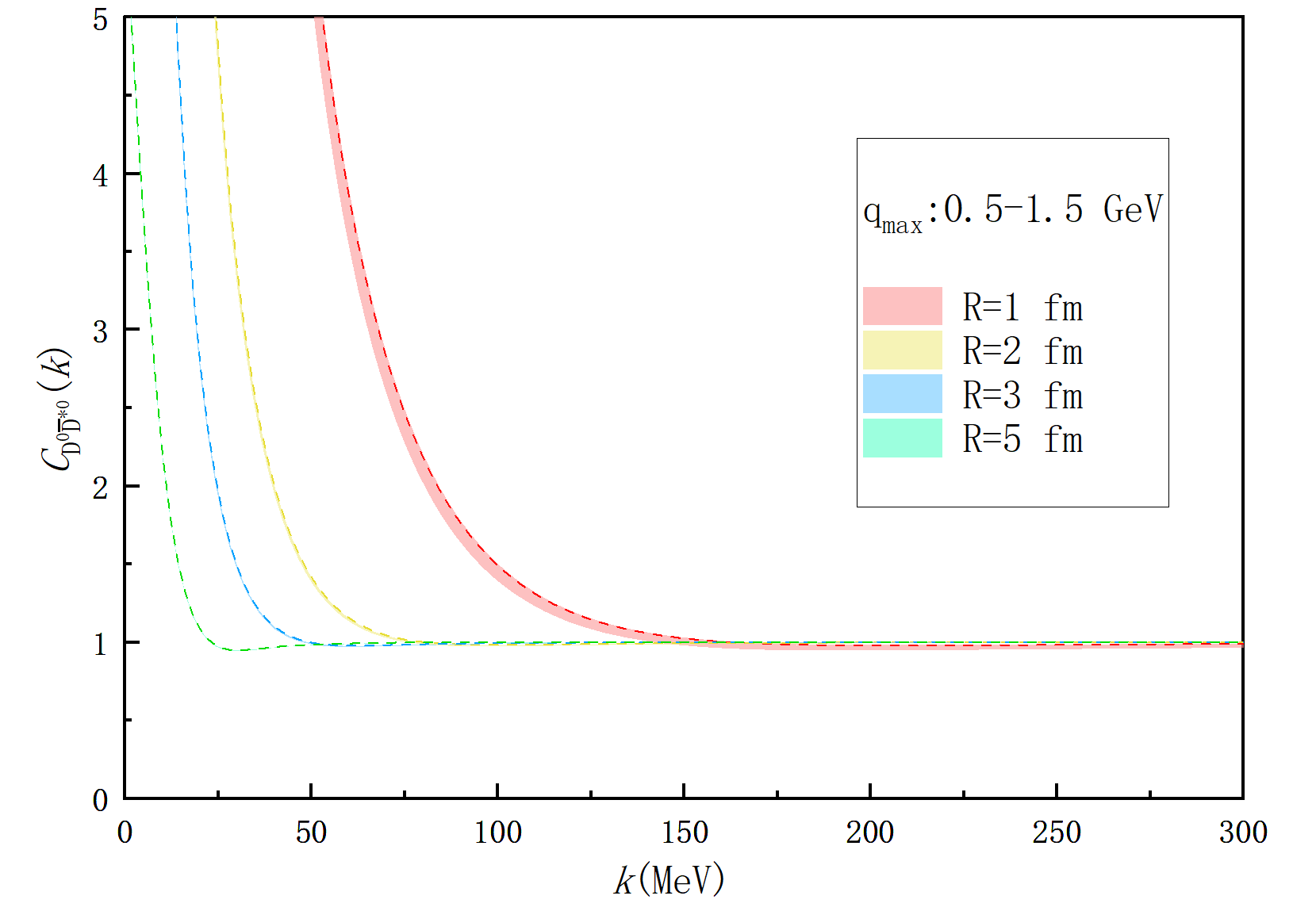} 
\caption{$D^0\bar{D}^{*0}$ CFs of the $X(3872)$ in the single-channel case without the dressing of a bare state. 
\label{3872 single}
}
\end{figure}

First, assuming the $X(3872)$ as a bound state of  $D^{0}\bar{D}^{*0}$,   we obtain  its  scattering length $a=20.2$ fm and effective range   $r_{0}=0.28$ fm. At this point, it seems to violate the Breit-Wigner causality, which demands $r_{0} <0$ for a zero-range interaction~\cite{Matuschek:2020gqe}. This is because we set the cutoff to $1$ GeV, which does not correspond to a $\delta$ potential in coordinate space. Subsequently, its corresponding CFs are shown in Fig.~\ref{3872 single}, which are consistent with the general feature of CFs for a weakly bound state~\cite{Liu:2023uly}. 

\begin{figure}
\includegraphics[width=0.45\textwidth]{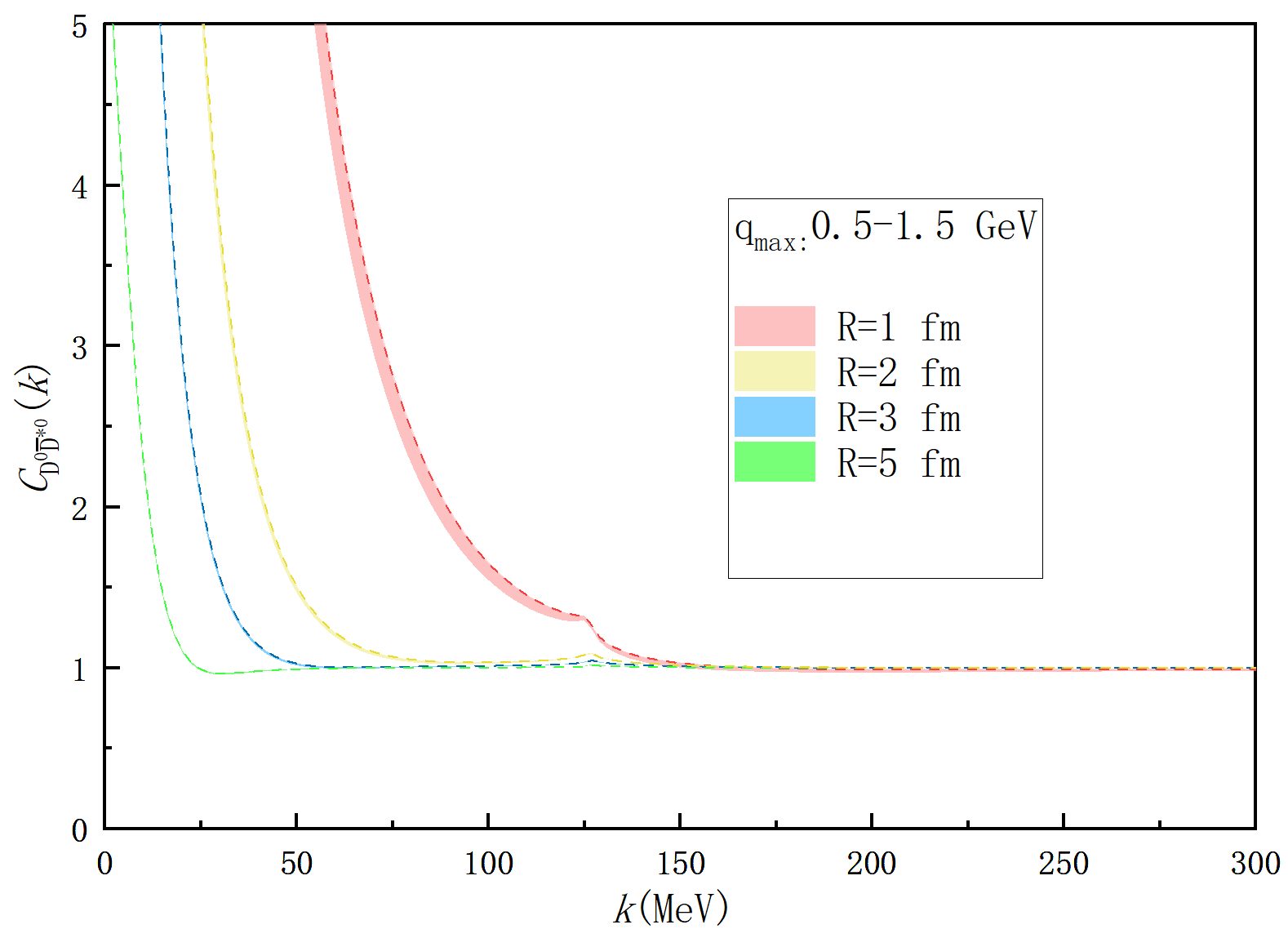} 
\caption{$D^0\bar{D}^{*0}$ CFs of the $X(3872)$ in the coupled-channel case without the dressing of a bare state. 
\label{3872}
}
\end{figure}

In addition to the neutral channel $D^{0}\bar{D}^{*0}$, the charge channel $D^{+}D^{*-}$ is incorporated into our analysis, resulting in a coupled-channel system. The coupled-channel contact EFT potential is presented in Ref.~\cite{Shen:2024npc}. Similarly, we obtain the 
 scattering length $a_{D^{0}\bar{D}^{*0}}\approx 19.58 $ fm and effective range $r=-1$ fm for the neutral channel, which is basically in line with the results of Ref.~\cite{Baru:2021ldu} and the LHCb result~\cite{LHCb:2020xds}. We can see that the coupled-channel effects affect the effective range more significantly than the scattering length.    
In Fig.~\ref{3872}, we present the corresponding $D^{0}\bar{D}^{*0}$ CFs, which show little discrepancy with those of the single-channel case in Fig.~\ref{3872 single}.  

\begin{figure}
\includegraphics[width=0.45\textwidth]{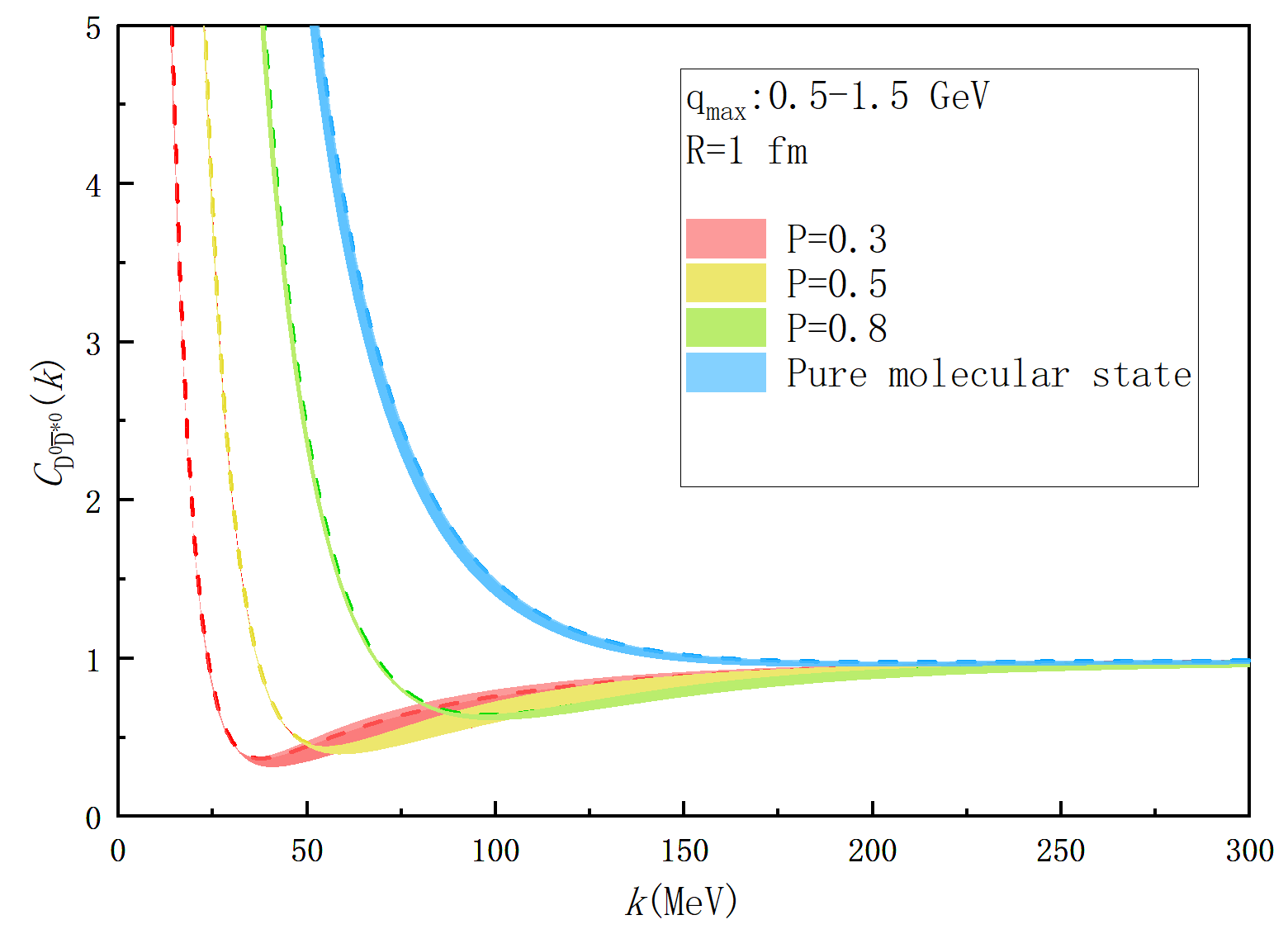} 
\caption{The same as Fig.~\ref{3872 single} but with the dressing of a bare state. 
\label{3872singlebare}
}
\end{figure}

\begin{figure}
\includegraphics[width=0.45\textwidth]{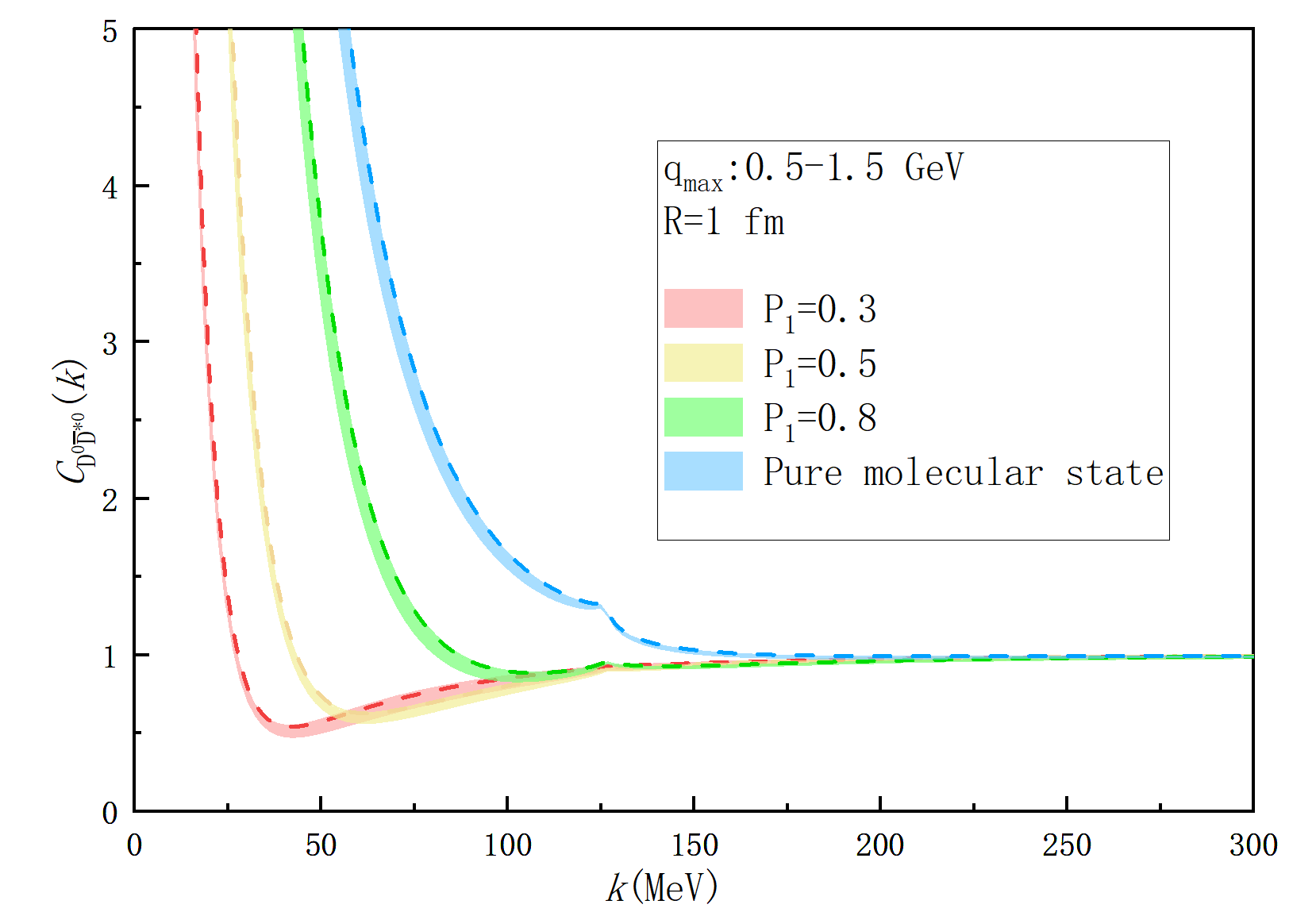} 
\caption{The same as Fig.~\ref{3872} but with the dressing of a bare state. 
\label{38722}
}
\end{figure}

In the following, we analyze the effect of including a bare state.  With the  molecular component of $X(3872)$ being  85\% ($P_{1}=0.8, P_{2}=0.05$), 
 we obtain the scattering length $a= 17.96$ fm and effective range $r_{1}= -4.72$ fm for the neutral channel, indicating that the bare state significantly affects both the scattering length and the effective range. Then  we   calculate the  $D^{0}\bar{D}^{*0}$ CFs for different compositeness  at $R=1$ fm, as shown in Fig.~\ref{3872singlebare} and Fig.~\ref{38722}. These results indicate that the dressing effect of the bare state significantly modifies the line shape of the CFs.  A clear dependence of the line shapes on the compositeness is observed. In particular, the CFs fall below unity within a certain range of compositeness, providing insight into the composition of this shallow bound state.

\section{Summary}

\begin{figure}[htpb]
  \centering
  \includegraphics[width=8.9cm]{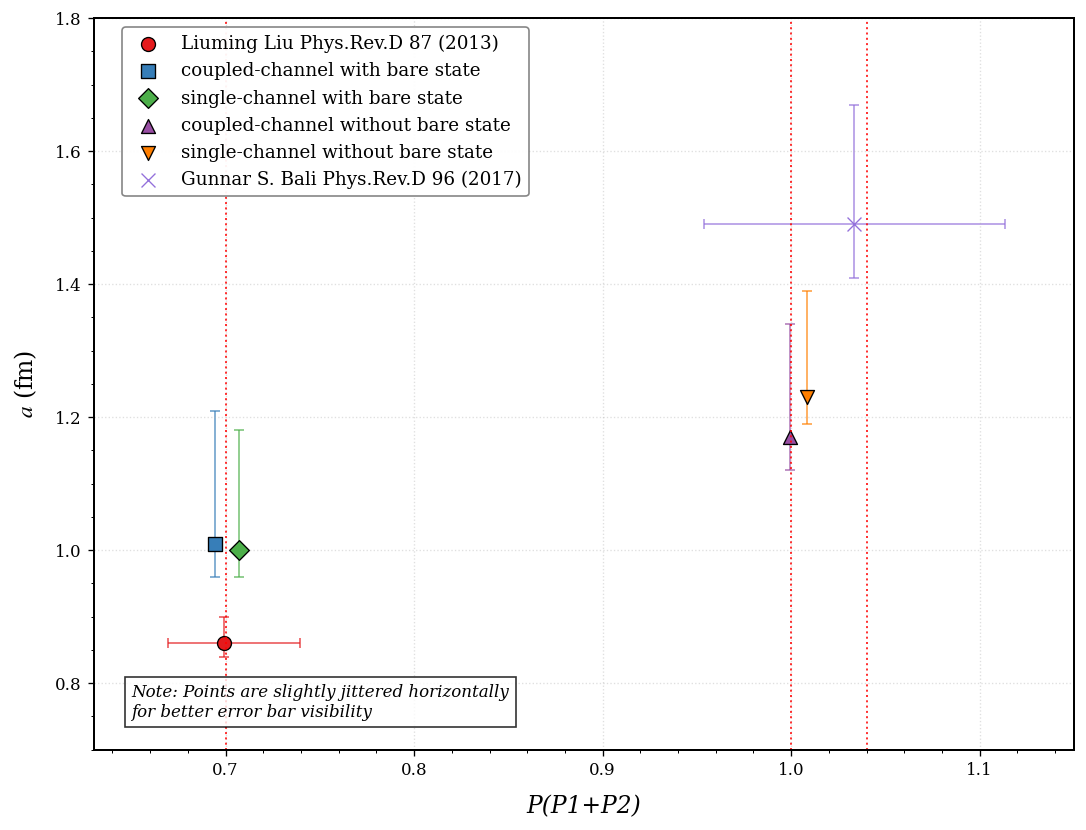}
  \caption{The scattering length $a$ for molecular proportions 70\% and 100 \% calculated by us and lattice QCD. The upper and lower limits of our results represent the case of $q_{max}$ $0.5$ GeV and $1$ GeV}
  \label{scatteringlength1}
\end{figure}

The discoveries of the exotic states $X(3872)$ and $D_{s0}^*(2317)$ motivated us to probe the nature of states beyond the conventional quark model. Recent theoretical and experimental investigations suggest that these states may possess substantial hadronic molecular components. In this work, we performed a systematic comparative study of  $X(3872)$ and $D_{s0}^*(2317)$ under two distinct scenarios: with and without bare-state contributions. Using a model-independent EFT approach, we extracted the hadron-hadron interactions and subsequently predicted the scattering lengths and  CFs for each scenario.

As shown in Fig.~\ref{scatteringlength1}, our calculations yielded a $DK$ scattering length of $1.2$ fm in the pure molecular scenario and $1.0$ fm including bare-state contributions. These values exhibit remarkable consistency with contemporary Lattice QCD predictions, indicating the validity of our theoretical framework.  Then, using a similar theoretical framework for hadron-hadron potentials, we predicted the relevant CFs.  Our results indicate that the lineshape of the $D^{0}K^{+}$ correlation functions is sensitive to the admixture effects from the coupled-channel $D^+K^0$ and the bare state. In addition, we find that the $D_s\eta$ channel plays only a minor role in determining the $D^{0}K^{+}$ correlation functions, consistent with observations from the scattering length.  It is worth noting that the $D^{0}K^{+}$ correlation function can probe the position of the bare state, if such a QCD bare state exists.

In addition,  we found that for $R=1$ fm, the lineshape of the correlation function changes most noticeably. We also verified that the weight of the individual components of the multi-channel wave function has a negligible effect on our results. We showed that the larger the proportion of molecular configurations, the larger the deviation of the CFs from $1$.
Notably, we resolved the inverse problem in CFs analysis by establishing a bijective relationship between compositeness and CFs. We demonstrated that the compositeness can be reliably extracted from the CFs, thereby establishing a reciprocal relationship between these two quantities. At last, we calculated the $D^{0}\bar{D}^{*0}$ CFs, revealing that the dressing effect of the bare state significantly modifies their line shapes. A clear dependence of the line shapes on the compositeness is observed. The CFs fall below unity within a certain range of compositeness, offering insight into the composition of this shallow-bound state.

\noindent {\it Acknowledgement.—}
This work is partly supported by the Fundamental and Interdisciplinary Disciplines Breakthrough Plan of the Ministry of Education of China-JYB2025XDXM204 and the National Natural Science Foundation of China under Grant No. W2543006 and No. 12435007. Ming-Zhu Liu acknowledges support from the National Natural Science Foundation of China under Grant No.12575086. Zhi-Wei Liu acknowledges support from the National Natural Science Foundation of China under Grant No.12405133, and Shenzhen 
Science and Technology Program under Grant No.ZDSYS20230626091501002.

\bibliography{reference}

\end{document}